# Remarks upon neutrino mixing hypothesis

**Viliam Pažma**[1] **and Julius Vanko**[2]

1/ Department of Theoretical Physics and Didactics of Physics
2/ Department of Nuclear Physics and Biophysics
Comenius University, Mlynska dolina F1, 84248 Bratislava, Slovakia
vanko@fmph.uniba.sk

**Abstract**

It is shown that various versions of the neutrino mixing hypothesis are in a contradiction with generally accepted facts and principles. There is also presented the possible alternative formulation of the neutrino oscillation theory and it is shown under what conditions this theory reproduces the known oscillation probability formula. In our approach (flavor) neutrinos are Dirac particles. In the case of Majorana neutrinos, or the nonrelativistic neutrinos (i.e. relic neutrinos) the problem could be more complicated.

**1. Introduction**

The neutrino mixing hypothesis is the cornerstone of the contemporary formulation of the neutrino oscillation theory (see e.g. [1-5]). In accordance with this hypothesis the states $|v_\alpha\rangle$, ($\alpha = e, \mu, \tau$) of the free flavor neutrinos $v_e$, $v_\mu$, $v_\tau$ are defined by means of the states $|v_i\rangle$, ($i = 1, 2, 3$) of the so-called massive free neutrinos $v_i$ by the relation

$$|v_\alpha\rangle = U_{\alpha i} |v_i\rangle, \qquad (1)$$

where $U_{\alpha i}$ are elements of an unitary matrix $U$ ($U$ is constant matrix) and the summation over repeated indices is assumed.

The massive free neutrino states $|v_i\rangle$ have the masses $M_i$, the helicity $(-\tfrac{1}{2})$ and they are eigenstates of the generator $H$ of the time development. The definition (1) is too formal and one can encounter its various versions in the literature :

i/
$$|v_\alpha; \vec{p}\rangle = U_{\alpha i} |v_i; \vec{p}, M_i\rangle. \qquad (2)$$

The equation (2) represents the so-called equal momentum case. It is well-known that (2) is not compatible with the Lorentz invariance (see e.q. [1] ). Namely if some observer $O$ registers $U_{\alpha i}|v_i, \vec{p}, M_i\rangle$ as the flavor state of $v_\alpha$ than another observer $O'$ who moves with the velocity $\vec{v} \neq 0$ with respect to $O$, will register the state

$$U_{\alpha j}|v_i; \vec{p}'_i, M_i\rangle \qquad (3)$$

where $\vec{p}'_i \neq \vec{p}'_j$ if $M_i \neq M_j$. Thus in accordance with (2) the observer $O'$ will not consider (3) as the state of flavor neutrino $v_\alpha$.

ii/
$$|v_\alpha\rangle = U_{\alpha i}|v_i; \vec{p}_i; M_i\rangle \qquad (4)$$

where

$$E = \sqrt{\vec{p}_i^2 + M_i^2} \qquad \text{for all } i = 1,2,3,$$

(the equal energy case (see e.g. [1])). The definition (4) is not also compatible with the Lorentz invariance. Moreover, it is evident that right-hand side of (4) is the eigen-state of $H$ so transitions to other states do not exist. In [1] there is solved the following problem : "How to take into account the different momenta of massive neutrinos in the derivations of the oscillation probability ?". Hence the author consider the following version of (1)

iii/
$$|v_\alpha\rangle = U_{\alpha i}|v_\alpha, \vec{p}_i, M_i\rangle \qquad (5)$$

This definition of $|v_\alpha\rangle$ hardly can be accepted. Let us consider two trinities $(\vec{p}_1, \vec{p}_2, \vec{p}_3)$, $(\vec{p}_1, \vec{p}_2, \vec{p}'_3 \neq \vec{p}_3)$. In the first case we have, e.g.,
$$|v_e\rangle = U_{ei}|v_i, \vec{p}_i, M_i\rangle$$

whereas for second it is
$$|v_\mu\rangle = \sum_{i=1}^{2} U_{\mu i}|v_i, \vec{p}_i, M_i\rangle + U_{\mu 3}|v_3, \vec{p}'_3, M_3\rangle.$$

Then
$$\langle v_e | v_\mu \rangle = \sum_{i=1}^{2} U^*_{ei} U_{\mu i} \neq 0.$$

Moreover, as the kernel of the considerations presented in [1] is the oscillation probability formula of the form

$$P_{v_\alpha \to v_\beta}(L, T) = \left| U^*_{\alpha k} e^{ip_k L - iE_k T} U_{\beta k} \right|^2 \qquad (6)$$

where $L$ is source-detectordistance and $T$ is time which passed from the production of $v_\alpha$ till the detection of $v_\beta$. This result evokes a certain suspicion. Namely for $T \to 0$ we get

$$P_{\nu_\alpha \to \nu_\beta}(L,0) = \left| U^*_{\alpha k} e^{ip_k L} U_{\beta k} \right|^2 \neq \delta_{\alpha\beta} \;,$$

whereas it is natural to expect that for $T = 0$ no transition $\nu_\alpha \to \nu_\beta \neq \nu_\alpha$ is possible. This simply means that considerations presented in [1] are not consistent. Hence, the neutrino mixing hypotheses (2) or (4) are reference frame dependent.

In [6] there was presented the reference frame independent definitions of the flavor states. It reads

$$|\nu_\alpha\rangle = U_{\alpha i} \left| \nu_i, \vec{p}_i, M_i \right\rangle \tag{7}$$

where

$$\frac{\vec{p}_i}{E_i} = \frac{\vec{p}_i}{\sqrt{\vec{p}_i^{\,2} + M_i^2}} = \vec{v} = const.$$

for all $i = 1, 2, 3$ (equal velocity case). Let us now consider the interaction of $\nu_\alpha$ with charged lepton $l_\alpha$. Considering the case when $l_\alpha$
gained the momentum $\vec{k}$. The initial state $\left| l_\alpha, \vec{p}, m_\alpha \right\rangle U_{\alpha i} \left| \nu_i, \vec{p}_i, M_i \right\rangle_{\vec{v}}$ will convert to the final state

$$\left| l_\alpha, \vec{p} + \vec{k}, m_\alpha \right\rangle A(t, \vec{p}, m_\alpha, \vec{p}_i, M_i, \vec{k}) \; U_{\alpha i} \left| \nu_i, \vec{p}_i - \vec{k}, M_i \right\rangle . \tag{8}$$

All $\vec{p}_i$ are parallel and if $\vec{k}$ is not parallel with $\vec{p}_i$ then $\vec{p}_i - \vec{k}$ $(i = 1,2,3)$ are not parallel and so (8) cannot be written in the form

$$\left| l_\alpha, \vec{p} + \vec{k}, m_\alpha \right\rangle \tilde{A} \; U_{\alpha i} \left| \nu_i, \vec{p}_i - \vec{k}, M_i \right\rangle_{\vec{v}'} .$$

Hence, if interaction $\nu_\alpha$ with $l_\alpha$ is defined by means of interactions $\nu_i$ with $l_\alpha$ then in the final state we shall not obtain the flavor neutrino $\nu_\alpha$ defined by (7). The previous remarks evoke the question : What version of the neutrino mixing hypothesis can be acceptable and in what sense ? In the next section we shall show that (2) (equal momentum case) can be accepted as good approximation in the case of ultra-relativistic neutrinos. Naturally, instead of (2) we have to write

$$|\nu_\alpha\rangle \approx U_{\alpha i} \left| \nu_i; \vec{p}, M_i \right\rangle .$$

## 2. Alternative approach to neutrino oscillations

In this section we shall regard $\nu_e, \nu_\mu, \nu_\tau$ as Dirac particles and formulate (phenomenological) the theory of their oscillations. We shall also show that this theory reproduces the results following from the standard one (based on (2)) in the region $\vec{p}^2 \gg$ squared mass of any neutrino.

Let us now define the (flavor) neutrino wave functions $\langle p| \nu_i \rangle$ (we work in the p-representation and instead of indices $\alpha = e, \mu, \tau$ we shall use the indices $i = 1, 2, 3$ ) as the eigenfunctions of the hamiltonian

$$H_0 = \vec{\alpha} \cdot \vec{p} + \beta M_d \qquad (9)$$

where $M_d = diag\,(m_1, m_2, m_3)$, $m_i$'s are masses of $\nu_i$ and the standard meaning of other symbols is assumed. In the standard representation of Dirac matrices and for $\vec{p} = (0, 0, p > 0)$ we can choose $\langle p|\nu_i\rangle$ as

$$\langle p|\nu_i\rangle = \frac{U_i}{\sqrt{2\varepsilon_i}}, \qquad (\varepsilon_i = \sqrt{p^2 + m_i^2}),$$

where for $i = 1, 2, 3$ we have

$$U_i = \begin{pmatrix} \delta_{1i}\, u_i \\ \delta_{2i}\, u_i \\ \delta_{3i}\, u_i \end{pmatrix}, \qquad u_i = \begin{pmatrix} \sqrt{\varepsilon_i + m_i}\ w \\ -\sqrt{\varepsilon_i - m_i}\ w \end{pmatrix},$$

$\sigma_3 w = -w$

These eigenfunctions correspond to positive eigenvalues of $H_0$ and the negative helicity and for $i = 4, 5, 6$

$$U_i = \begin{pmatrix} \delta_{4i}\, u_i \\ \delta_{5i}\, u_i \\ \delta_{6i}\, u_i \end{pmatrix}, \qquad u_i = \begin{pmatrix} \sqrt{\varepsilon_i - m_i}\ w \\ \sqrt{\varepsilon_i + m_i}\ w \end{pmatrix},$$

$m_i = m_{i-3}, \quad \varepsilon_i = \varepsilon_{i-3}$

These eigenfunctions correspond to negative eigenvalues of $H_0$ and the negative helicity.

Now if
$$H = H_0 + \beta M' = \vec{\alpha} \cdot \vec{p} + \beta M$$

where $(M')^+ = M'$ and $M'_{ii} = 0$ for all $i$, is the generator of the time-development then the transitions $v_i \to v_j$ will occur in the theory. (As far as we know this approach was considered in [5,6]). Now the question arises. Is there any point of contact between this theory and the standard one ? In the next we shall show that the reply is yes.

Let $C$ be matrix satisfying $C M C^+ = diag(M_1, M_2, M_3)$ where $M_i$ are eigenvalues of $M$. The eigenfunctions of $H$ corresponding to negative helicity we can choose in the form

$$\langle p|v'_i\rangle = \frac{V_i}{\sqrt{2E_i}} = \frac{C^+ V'_i}{\sqrt{2E_i}}, \qquad (E_i = \sqrt{p^2 + M_i^2}),$$

where for $i = 1, 2, 3$

$$V'_i = \begin{pmatrix} \delta_{1i} u'_i \\ \delta_{2i} u'_i \\ \delta_{3i} u'_i \end{pmatrix}, \qquad u'_i = \begin{pmatrix} \sqrt{E_i + M_i} \; w \\ -\sqrt{E_i - M_i} \; w \end{pmatrix}$$

(they correspond to positive eigenvalues of $H$) and for $i = 4, 5, 6$
($E_4 = E_1, E_5 = E_2, E_6 = E_3$)

$$V'_i = \begin{pmatrix} \delta_{4i} u'_i \\ \delta_{5i} u'_i \\ \delta_{6i} u'_i \end{pmatrix}, \qquad u'_i = \begin{pmatrix} \sqrt{E_i - M_i} \; w \\ \sqrt{E_i + M_i} \; w \end{pmatrix}$$

(they correspond to negative eigenvalues of $H$). Because it holds

$$\frac{U_i^+ U_j}{2\sqrt{\varepsilon_i \varepsilon_j}} = \frac{V_i^+ V_j}{2\sqrt{E_i E_j}} = \delta_{ij},$$

$$\sum_{i=1}^{6} \frac{U_i U_i^+}{2\varepsilon_i} = \sum_{i=1}^{6} \frac{V_i V_i^+}{2E_i} = diag\,(010101010101).$$

Then we can write

$$\frac{U_i}{\sqrt{2\varepsilon_i}} = T_{ij}\frac{V_j}{\sqrt{2E_j}} \qquad (9)$$

or

$$T_{ij} = \frac{V_j^+ U_i}{2\sqrt{E_j \varepsilon_i}}. \qquad (10)$$

Putting

$$T = \begin{pmatrix} T^{(1)} & T^{(2)} \\ T^{(3)} & T^{(4)} \end{pmatrix}$$

where $T^{(\alpha)}$ are 3 x 3 matrices then from (11) we get

$$T_{ij}^{(1)} = C_{ji}\frac{1}{2\sqrt{E_j \varepsilon_i}}\left(\sqrt{(\varepsilon_i + m_i)(E_j + M_j)} + \sqrt{(\varepsilon_i - m_i)(E_j - M_j)}\right),$$

$$T_{ij}^{(2)} = C_{ji}\frac{1}{\sqrt{E_j \varepsilon_i}}\left(\sqrt{(\varepsilon_i + m_i)(E_j - M_j)} + \sqrt{(\varepsilon_i - m_i)(E_j + M_j)}\right),$$

(similary for $T^{(3)}$ and $T^{(4)}$). Thus, in the region $\vec{p}^2 \gg m_i^2, M_i^2$ (for all $i = 1,2,3$) we can write

$$T^{(1)} \approx C^T \approx T^{(4)}, \qquad T^{(2)} \approx T^{(3)} \approx 0.$$

Hence in the ultrarelativistic case the equation (10) can be approximated by

$$\frac{U_i}{\sqrt{2\varepsilon_i}} \approx (C^T)_{ij}\frac{V_j}{\sqrt{2E_j}}, \qquad (i,j = 1,2,3), \qquad (11)$$

or in more familiar form the last equation can be written as

$$|v_i\rangle \approx (C^T)_{ij}|v_j'\rangle. \qquad (12)$$

where $|v_i\rangle$ are states of flavor neutrinos and $|v_i'\rangle$ can be interpreted as so-called massive neutrinos states (eigenstates of the generator $H$ of the time-development). We remark that for $\vec{p} = 0$ we get $|v_i\rangle = (C^T)_{ij}|v_j'\rangle$.

## 4. Concluding remarks

Within the framework of the presented considerations the neutrino oscillations probability formula can be derived by the standard way. Having at time $t = 0$ the state $|v_i\rangle_0$ (corresponding to the momentum $\vec{p}$) then the time-development of this state is given by

$$|v\rangle_t = e^{-iHt} |v_i\rangle_0 = \sum_{j=1}^{6} \sum_{k=1}^{6} T_{ij} \, e^{-iE_j t} \, T^+_{jk} \, |v_k\rangle_0 \, .$$

Hence, the amplitude $A$ of the transition $v_i \to v_k$ at time $t$ is equal to

$$A(v_i \to v_k; t) = \sum_{j=1}^{6} T_{ij} \, e^{-iE_j t} \, T^+_{jk} \, .$$

In the region $\vec{p}^2 \gg m_i^2, M_i^2$ (for all $i = 1, 2, 3$) the last formula can be approximated by

$$A(v_i \to v_k; t) \approx \sum_{j=1}^{3} \left( U_{ij} \, e^{-\frac{iM_j}{2p} t} \, U^+_{jk} \right) e^{-ipt}$$

where $U = C^T$ and $C$ is given by the equation

$$CMC^+ = diag \, (M_1, M_2, M_3) \, .$$

Our previous considerations generate several questions and we want to mention some of them at least. First of all, what does $H$ describe actually ? Does it describe the oscillations of $v_e, v_\mu, v_\tau$ or a triplet of free Dirac particles with masses $M_1, M_2, M_3$ ? . If $M_i \geq 0$ then evidently $H$ describes a triplet of free particles with masses $M_i$ and and oscillations in our approach are something artificial (The case when $M_i$ is negative requires a more detailed study.).

Let us for a while consider that the transitions $v_i \to v_j$ are caused by the interaction of $v_i$ with some fields $\varphi_{ij}$ ($\varphi_{ii} = 0$) and the interaction term is $\sim (\overline{v}_i \, \varphi_{ij} \, v_j + h.c.)$ . If we shall replace $\varphi_{ij}$ by some constants, say $M'_{ij}$ then we obtain (semi)phenomenological theory outlined above. This possibility that neutrino oscillations are caused by the interaction of neutrinos with some field seems to be very tempting.

We are not sure if all these questions are meaningful but we feel that a detailed analysis of basic postulates of the theory in question is entitled and desirable.